\documentstyle[prl,aps,floats]{revtex}
\input epsf
\begin{document}
\def\met{$E \! \! \! \! / _T \:$} % missing ET
\lefthyphenmin=2
\righthyphenmin=3
\title{ Measurement of 
        the High-Mass Drell-Yan Cross Section and Limits on 
          Quark-Electron Compositeness Scales}

\author{                                                                      
%% names begin here                                                           
B.~Abbott,$^{40}$                                                             
M.~Abolins,$^{37}$                                                            
V.~Abramov,$^{15}$                                                            
B.S.~Acharya,$^{8}$                                                           
I.~Adam,$^{39}$                                                               
D.L.~Adams,$^{49}$                                                            
M.~Adams,$^{24}$                                                              
S.~Ahn,$^{23}$                                                                
H.~Aihara,$^{17}$                                                             
G.A.~Alves,$^{2}$
% ...
\\                                                                            
\vskip 0.70cm                                                                 
\centerline{(D\O\ Collaboration)}                                             
\vskip 0.70cm                                                                 
}                                                                             
\address{                                                                     
\centerline{$^{1}$Universidad de Buenos Aires, Buenos Aires, Argentina}       
\centerline{$^{2}$LAFEX, Centro Brasileiro de Pesquisas F{\'\i}sicas,         
                  Rio de Janeiro, Brazil}                                     
\centerline{$^{3}$Universidade do Estado do Rio de Janeiro,                   
                  Rio de Janeiro, Brazil}                                     
\centerline{$^{4}$Universidad de los Andes, Bogot\'{a}, Colombia}             
\centerline{$^{5}$DAPNIA/Service de Physique des Particules, CEA, Saclay,     
                  France}                                                     
\centerline{$^{6}$Panjab University, Chandigarh, India}                       
\centerline{$^{7}$Delhi University, Delhi, India}                             
\centerline{$^{8}$Tata Institute of Fundamental Research, Mumbai, India}      
\centerline{$^{9}$Kyungsung University, Pusan, Korea}                         
\centerline{$^{10}$Seoul National University, Seoul, Korea}                   
\centerline{$^{11}$CINVESTAV, Mexico City, Mexico}                            
\centerline{$^{12}$Institute of Nuclear Physics, Krak\'ow, Poland}            
\centerline{$^{13}$Institute for Theoretical and Experimental Physics,        
                   Moscow, Russia}                                            
\centerline{$^{14}$Moscow State University, Moscow, Russia}                   
\centerline{$^{15}$Institute for High Energy Physics, Protvino, Russia}       
\centerline{$^{16}$University of Arizona, Tucson, Arizona 85721}              
\centerline{$^{17}$Lawrence Berkeley National Laboratory and University of    
                   California, Berkeley, California 94720}                    
\centerline{$^{18}$University of California, Davis, California 95616}         
\centerline{$^{19}$University of California, Irvine, California 92697}        
\centerline{$^{20}$University of California, Riverside, California 92521}     
\centerline{$^{21}$Florida State University, Tallahassee, Florida 32306}      
\centerline{$^{22}$University of Hawaii, Honolulu, Hawaii 96822}              
\centerline{$^{23}$Fermi National Accelerator Laboratory, Batavia,            
                   Illinois 60510}                                            
\centerline{$^{24}$University of Illinois at Chicago, Chicago,                
                   Illinois 60607}                                            
\centerline{$^{25}$Northern Illinois University, DeKalb, Illinois 60115}      
\centerline{$^{26}$Northwestern University, Evanston, Illinois 60208}         
\centerline{$^{27}$Indiana University, Bloomington, Indiana 47405}            
\centerline{$^{28}$University of Notre Dame, Notre Dame, Indiana 46556}       
\centerline{$^{29}$Purdue University, West Lafayette, Indiana 47907}          
\centerline{$^{30}$Iowa State University, Ames, Iowa 50011}                   
\centerline{$^{31}$University of Kansas, Lawrence, Kansas 66045}              
\centerline{$^{32}$Louisiana Tech University, Ruston, Louisiana 71272}        
\centerline{$^{33}$University of Maryland, College Park, Maryland 20742}      
\centerline{$^{34}$Boston University, Boston, Massachusetts 02215}            
\centerline{$^{35}$Northeastern University, Boston, Massachusetts 02115}      
\centerline{$^{36}$University of Michigan, Ann Arbor, Michigan 48109}         
\centerline{$^{37}$Michigan State University, East Lansing, Michigan 48824}   
\centerline{$^{38}$University of Nebraska, Lincoln, Nebraska 68588}           
\centerline{$^{39}$Columbia University, New York, New York 10027}             
\centerline{$^{40}$New York University, New York, New York 10003}             
\centerline{$^{41}$University of Rochester, Rochester, New York 14627}        
\centerline{$^{42}$State University of New York, Stony Brook,                 
                   New York 11794}                                            
\centerline{$^{43}$Brookhaven National Laboratory, Upton, New York 11973}     
\centerline{$^{44}$Langston University, Langston, Oklahoma 73050}             
\centerline{$^{45}$University of Oklahoma, Norman, Oklahoma 73019}            
\centerline{$^{46}$Brown University, Providence, Rhode Island 02912}          
\centerline{$^{47}$University of Texas, Arlington, Texas 76019}               
\centerline{$^{48}$Texas A\&M University, College Station, Texas 77843}       
\centerline{$^{49}$Rice University, Houston, Texas 77005}                     
}       

\date{\today}

\maketitle

\begin{abstract}

 We present a measurement  of the Drell-Yan cross section at high
 dielectron
  invariant mass using 120 pb$^{-1}$ of data collected in 
 {\mbox{$p\bar p$}}\ collisions at {\mbox{$\sqrt{s}$ =\ 1.8\ TeV }}
 by the D\O\ 
 collaboration during 1992--96.  
 No deviation from  standard model expectations is observed. 
 We use the data to set limits on the energy scale of quark-electron
 compositeness with common constituents. 
 The 95\% confidence level lower limits on the compositeness scale vary 
 between 3.3 TeV and 6.1 TeV depending on the assumed form of the
 effective contact interaction. 

\end{abstract}

\vspace*{0.2cm}

\hspace*{1.5cm}
PACS numbers: 12.60.Rc, 13.85.Qk, 13.85.Rm

\twocolumn

 In $p \bar{p}$ collisions, dielectron pairs can be produced through the
 Drell-Yan 
 process \cite{DrellYan} over a large range of  available partonic center 
 of mass energies. 
 In the standard model (SM), 
 the process occurs to first order via quark-antiquark annihilation
 into a virtual photon or 
 $Z$ boson. 
 If quarks and leptons are composite with common constituents, the
 interaction of  
 these constituents would likely be manifested through an 
 effective four fermion contact interaction at energies below the 
 compositeness scale. 
  We consider a general contact-interaction 
 Lagrangian \cite{llmodel,Taekoon}
 of the form
\begin{flushleft}
\begin{eqnarray}
 {\cal L} &=& \frac {4 \pi} {\Lambda^2}
 [   \eta_{LL}(\bar{q}_L \gamma^{\mu} q_L) (\bar{e}_L \gamma_{\mu} e_L) 
 + \eta_{LR} (\bar{q}_L \gamma^{\mu} q_L) (\bar{e}_R \gamma_{\mu} e_R)
 \nonumber \\ 
 &+& \eta_{RL}(\bar{q}_R \gamma^{\mu} q_R) (\bar{e}_L \gamma_{\mu} e_L) 
 + \eta_{RR}(\bar{q}_R \gamma^{\mu} q_R) (\bar{e}_R \gamma_{\mu} e_R) ]
 \nonumber 
\label{lagrangian}
\end{eqnarray} 
\end{flushleft}
 where $q$=$(u,d)$ represents the first generation quarks, $\Lambda$ is the 
 compositeness scale,  $\eta_{ij}
 = \pm 1$, and 
 $L$ $(R)$ denotes the left (right) helicity 
 projection. 
  The addition of this contact term to the SM Lagrangian 
  modifies the dominant $\gamma/Z$ boson matrix element, with the
 strongest effects  at high dielectron invariant mass.
 Composite quarks and 
 electrons have been proposed
 as a possible explanation of the high-$Q^2$ anomaly at HERA
 \cite{HERA}. 
 Previous results  on quark-electron compositeness 
  set
 lower limits on the 
 energy scale $\Lambda$ in the range of 2.5--5.2 TeV \cite{cdfprl}
  and 2.1--3.5 TeV \cite{opal}.
 In this Letter, we report the 
 measurement of the Drell-Yan cross section at high mass 
 and set the most stringent limits to date
 on the quark-electron compositeness scale. 

 The results presented here used 120 pb$^{-1}$ of data 
  collected in
  $p\bar{p}$ collisions 
 at  $\sqrt{s}=1.8$ TeV by the D$\O$ detector \cite{D0detector} during the
 1992--1996  run at the Fermilab Tevatron.
 The  detector consists of a tracking system, 
 a highly linear, granular and stable
 uranium/liquid-argon calorimeter,  and a muon spectrometer. 
 Electron candidates are accepted in the 
 pseudorapidity range of $\mid$$\eta$$\mid$ $<1.1$ for 
 electrons detected in the central calorimeter (CC) 
 and $1.5<$ $\mid$$\eta$$\mid$ $<2.5$ for electrons detected in the forward
 calorimeters (EC), where $\eta = -{\rm log \; tan} (\theta/2)$ and $\theta$
 is the polar angle with respect to the beam axis.
 CC electrons within 0.0098 radians in  azimuth 
 of any calorimeter module edge are removed
  to ensure uniform calorimeter response. 
 At least two electrons are required
 to have transverse energy $E_T >  20 $ GeV at the trigger level and 
 $E_T > 25 $ GeV offline.  
 The two highest-$E_T$ electrons in the event are selected for analysis. 

  Offline, a ``loose'' electron must satisfy three requirements: 
 (i) the electron must deposit at least 95\% of its energy in the 21 
 X$_0$ electromagnetic calorimeter, 
 (ii) the transverse and longitudinal shower shapes must  be consistent with 
 those expected for an electron, and  
 (iii) the electron must be isolated in energy in a cone of radius 
 $R = \sqrt{\Delta \eta^2 + \Delta \phi^2} = 0.4$, 
 such that $\frac{E_{tot}(R=0.4)
 - E_{EM}(R=0.2)}{E_{EM}(R=0.2)}<0.15$, where $E_{tot}$ and $E_{EM}$ are the
 total and EM calorimeter energies respectively. 
 A ``tight'' electron is additionally required 
  to have a matching track in the drift chambers. 
  In this analysis, any forward electron is required to be ``tight'' and at
 least one member of each electron pair must be ``tight.''

  The detector acceptance for dielectron events is defined as the
 fraction of produced events in which 
 both electrons pass our kinematic and fiducial cuts. 
 To calculate the acceptance, Drell-Yan events are generated using {\small
 PYTHIA} 
 \cite{pythia}.
   The parton showering parameters in 
 {\small PYTHIA} were tuned for good kinematic modeling of the data
 using the distribution of
 the $Z$ boson transverse momentum  
 ($p_T$)  observed at  D\O~$\! \!$. 
 The detector response is simulated using a parameterized
 Monte Carlo program \cite{Wmassprd}. The sampling and noise terms in the
 electron energy resolution are 
 derived  from test beam data and the calorimeter pedestal distribution in
 $W \rightarrow e \nu$ collider data, respectively. 
 The constant term is constrained by the observed width 
 of the  $Z \rightarrow ee$ mass peak. 
 The known $Z$ boson 
 mass is used to set the electromagnetic energy
 scale.

 The acceptance, calculated using the
 Drell-Yan Monte Carlo model and 
 MRS(A$^\prime$) \cite{mrsa} 
 parton distribution functions, is $\approx$ 53\% and
 does not depend strongly
 on mass above $m_{ee}=250$ GeV/$c^2$. This makes the analysis relatively
 model-independent. 
 The systematic uncertainty on the acceptance due to
 the production model is estimated to be $1.5\%$. 
 The effect of energy smearing,  included in the
 acceptance calculation, is small because the energy
 resolution (15\%/$\sqrt{E \; {\rm (GeV)}}$ $\oplus$ 1\%) is much smaller than
 the bin width at high mass.   

 The electron trigger and offline selection efficiency  is determined 
 using $ Z \rightarrow ee$ data. One of the electrons is required to 
 satisfy the  
 ``tight'' selection criteria. The second electron then 
 provides an unbiased sample 
 to measure the efficiencies. Background subtraction is performed using the
 sidebands of the $Z$ boson mass distribution. The 
 trigger is  found
 to be  fully efficient for high mass dielectrons ($m_{ee}>120$ GeV/$c^2$).
 The single-electron efficiency for the CC ``loose'' selection 
 criteria is (92.9$\pm$0.7)\%  and for the ``tight'' selection criteria is 
 (74.1$\pm$0.6)\%. The efficiency for EC ``tight'' selection criteria is 
 (52.6$\pm$1.0)\%.  The dependence of dielectron selection efficiency on 
 invariant mass was
 studied using a detailed {\small GEANT}-based Monte Carlo \cite{geant}
 simulation of dielectron events. 
\begin{figure}[tbhp]
\epsfxsize=3.0in
\centerline{\epsfbox{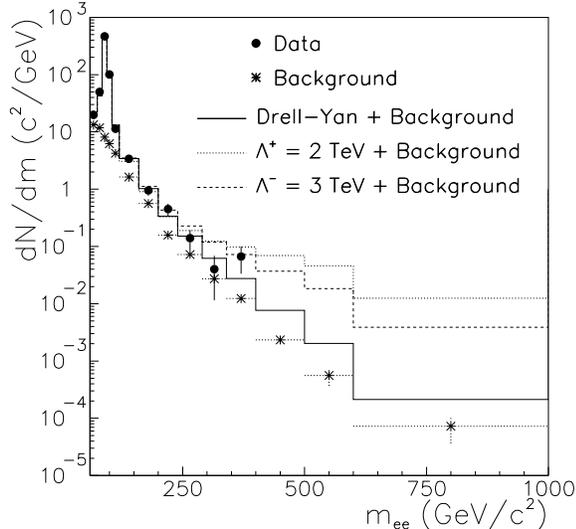}}
\caption { Mass distribution $dN/dm$ for dielectron data. 
 The corresponding distributions for expectations from Drell-Yan 
 and Drell-Yan $+$ contact term process (both including all other
 backgrounds) 
 are also shown. 
 The effects of kinematic
 and fiducial cuts, dielectron identification efficiency, and
  detector smearing are folded into the predictions, and the 
 predictions are normalized to the total luminosity. 
 Error bars indicate only statistical uncertainties. There are no events
 with $m_{ee} > 400$ GeV$/c^2$. }
\label{dyplot}
\end{figure}
 The Monte Carlo events were embedded in unbiased  data events
 to simulate the effects of multiple
 interactions and detector noise, and then reconstructed. 
 No dependence
 of  the selection efficiency on  dielectron mass was found. 

 The most important sources of background to $ p \bar{p} \rightarrow ee + X$ 
 are QCD multijet events with two jets misidentified as electrons and
 direct-photon 
 events where both the photon and a jet are misidentified as electrons.  
 Jets with a leading $\pi^0$ or $\eta$ may
  produce an isolated and energetic
 photon that  
 passes the ``loose'' or ``tight'' electron selection criteria, 
 depending on the  presence of  an  
 associated track.
 Using multijet and photon-jet data samples,  the probability for 
 misidentifying a jet as an electron is measured as a function of jet $E_T$.  
 A CC jet with  $E_T$ = 100 GeV is misidentified as a 
 ``tight'' 
 electron with a probability of  0.8$\times 10^{-3}$ and as a 
  ``loose'' electron with a probability of 1.8$\times 10^{-3}$. 
 An EC jet with  $E_T$ = 100 GeV is misidentified as a 
 ``tight'' 
 electron with a probability of  1.0$\times 10^{-3}$. 
 The 25\% uncertainty in these probabilities is dominated by the uncertainty
 in the direct-photon fraction of the fake electron sample.  
 Estimated backgrounds to the dielectron sample 
 from multijet, photon-jet, and $W$+jet sources were 
 calculated independently as a function of mass
  by weighting the total number of
 dijet, photon-jet, and 
 $W$+jet pairs in any given mass bin by the appropriate misidentification 
 probability and  sample luminosities. 

\begin{table} [tbhp]
\caption{The observed number of events $N$, total detection efficiency, 
 expected total background, and the dielectron production
 cross section in the given mass bins. In the three highest mass bins, we
 quote the 95\% (84\%) 
 CL upper limits on the cross section. The last column gives
 the value $\tilde {m}$ of mass at which, for a NNLO SM calculation, 
  $d \sigma^{\rm th} / dm$ equals $\sigma ^{\rm th} / \Delta m$. Here $\sigma
 ^{\rm th}$ denotes the total theoretical
  cross section in the bin and $\Delta m$ 
 denotes  the bin width.}
\medskip
\begin{center}
\begin{tabular}{cccccc}
$m_{ee}$ bin & $N$ & Total & Expected  &  ${\sigma}$ & $\tilde {m}$ \\ 
(GeV$/c^2$)&     & Efficiency & Background  & (pb) & (GeV$/c^2$)\\
\hline 
120--160&136& 0.32$\pm$0.01&64.0$\pm$10.0  & $1.93^{+0.43}_{-0.44}$ & 135 \\
160--200&38& 0.34$\pm$0.01&22.0$\pm$3.5  & $0.49^{+0.16}_{-0.18}$ & 177 \\
200--240&18& 0.36$\pm$0.01&6.34$\pm$0.96 & $0.28^{+0.09}_{-0.10}$ & 218 \\
240--290&7 & 0.37$\pm$0.01&3.61$\pm$0.56 & $0.066^{+0.052}_{-0.058}$ & 262 \\
290--340&2 & 0.38$\pm$0.01&1.37$\pm$0.23 & $0.033^{+0.032}_{-0.030}$ & 312 \\
340--400&4 & 0.39$\pm$0.01&0.75$\pm$0.13 & $0.057^{+0.042}_{-0.047}$ & 367 \\
400--500&0 & 0.40$\pm$0.01&0.23$\pm$0.04& $<$0.063 (0.039)&  443 \\
500--600&0 & 0.41$\pm$0.01&0.06$\pm$0.02& $<$0.060 (0.037)&  544 \\
600--1000&0& 0.42$\pm$0.01&0.03$\pm$0.01& $<$0.058 (0.035)& 729 \\
\end{tabular}
\end{center}
\label{crosstable}
\end{table}

 In addition to misidentification backgrounds, other
 high $p_T$ 
 processes  contribute to dielectron final states. 
 We use {\small PYTHIA} Monte Carlo events, 
 passed through the parameterized detector simulation, to estimate 
 these backgrounds. Since electrons and photons will pass the
 ``loose'' electron selection cuts,
 we evaluated five  possible background
 processes, $W\gamma \rightarrow e  \nu \gamma$, 
 $Z\gamma \rightarrow ee \gamma$, 
 $ t\bar{t} \rightarrow eeX$, $WW \rightarrow eeX$ and 
 $ \gamma^*/Z \rightarrow \tau \tau \rightarrow eeX $, and found them to 
 contribute less than 10\% of the total background. 

 Figure~\ref{dyplot} displays the 
 differential mass  distribution $dN/dm$ 
 for dielectron data. It also shows the 
 expectation from the Drell-Yan 
 process, obtained by processing {\small PYTHIA}-generated events through the 
 parametric detector simulation, and includes contributions from summed 
 background, which is also shown
 separately. 
 Our data 
 show  no significant discrepancy from  SM expectations. 

\begin{table*}[tbhp]
\caption{95\% CL lower limit on energy scale of compositeness $\Lambda$ in
 TeV for 
different contact interaction models. The superscript on $\Lambda$ indicates
 the sign of $\eta_{ij}$, which governs the nature of the interference 
 (negative sign for constructive interference) 
 between the contact interaction and the SM Lagrangian. }
\medskip
\begin{center}
\begin{tabular}{c|c|c|c|c|c|c|c|c|c|c}
    &$LL$&$LR$&$RL$&$RR$&$LL+RR$&$LR+RL$&$LL-LR$&$RL-RR$&$VV$&$AA$ \\
\hline 
$\Lambda^{+}$ (TeV)
& 3.3 & 3.4 & 3.3 & 3.3 &  4.2  &  3.9  &  3.9  &  4.0 & 4.9 & 4.7 \\
\hline
$\Lambda^{-}$ (TeV)
& 4.2 & 3.6 & 3.7 & 4.0 &  5.1  &  4.4  &  4.5  &  4.3 & 6.1 & 5.5 \\
\end{tabular}
\end{center}
\label{lamda_model_limit}
\end{table*} 

\begin{figure}[tbhp]
\epsfxsize=3.0in
\centerline{\epsfbox{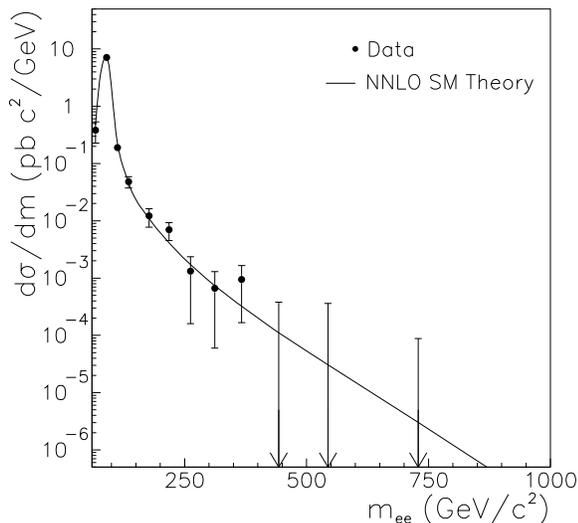}}
\caption {The differential inclusive dielectron production cross section.
  The 68\% uncertainty intervals are shown for the data points. The 
 last three bins, which have no events, show the 
 84\% CL upper limit on the cross section corresponding to the upper end
 of the error bars in the preceeding bins. 
 Also shown is the prediction
 of the  SM at NNLO.
}
\label{dycrossmeasure}
\end{figure}

  The measurement of the inclusive dielectron
  cross section is performed
  in independent mass bins using a Bayesian \cite{ETjaynes} technique. 
 In each bin $k$, we determine  
 the posterior probability density $P({\sigma^k}{\mid}{N_o^k})$ 
  for the  cross section $\sigma^k$, given the
 observed number of events $N_o^k$. 
 The expected number of events in the 
 $k^{\rm th}$ mass bin is given by 
 $N^k = b^k + {\cal L}{\epsilon^k} {\sigma^k}$, 
 where $b^k$ is the expected background, ${\cal L}$ is the  luminosity, 
 ${\epsilon^k}$ is the total 
 signal efficiency (including acceptance, selection efficiency, and smearing
 correction), and ${\sigma^k}$
 is the total cross section in that bin. 
 The posterior probability density for the cross section $\sigma^k$ 
 is 
\begin{eqnarray}
P({\sigma^k}{\mid}{N_o^k})=
\frac{1}{A}
\int db \; d{\cal L} \; {d\epsilon}
\frac {{e^{-{N^k}}}{N^k}^{N_o^k}} {{N_o^k}!}
P({b^k},
{\cal L},
{\epsilon^k})
P(\sigma^k), \nonumber
\label{post_prob_sigma}
\end{eqnarray} 
 where $A$ is the normalization. 
 The prior probability density
  $P({b},{\cal L},{\epsilon})$ is taken to be a product of 
 independent Gaussian distributions 
 in $b$, ${\cal L}$ and ${\epsilon}$, with the measured value in each bin
 defining the mean and
 the uncertainty defining the width. 
 The prior distribution $P({\sigma^k})$ in any bin is chosen to be 
  uniform in $\sigma$. 
 The measured value of the cross section for each bin is taken to be the 
 mode  of  
 the posterior probability density (maximum likelihood estimate). 
 The  interval of minimum width  containing  68\% of the area 
  defines
 the uncertainty on the cross section. 
 Table~\ref{crosstable} shows the observed number of 
  events, the product of detector acceptance and
 efficiency, and the expected background for 
 dielectron events. The second-to-last 
 column  shows the 
 measured dielectron cross section and the  associated uncertainty, 
  dominated by event 
 statistics in the high-mass bins. In bins with no observed events, we quote
 the 95\% and 84\% 
 confidence level (CL) upper limits on the cross section, defined
 by 
  $ \int_{0}^{\sigma} P({\sigma^\prime}{\mid}{N_o}) d \sigma^\prime=0.95$ 
 (0.84). 
The measured differential 
 cross sections $d \sigma / d m$
 are compared with predictions of a next-to-next-to-leading 
 order (NNLO) SM
 calculation \cite{dynnlo}
 in Fig.~\ref{dycrossmeasure}. We find no significant 
 deviation between the measurement and  theory. 

 To set limits on the compositeness scale $\Lambda$, 
 we calculate the cross section for the Drell-Yan $+$ contact term process 
  by 
 including terms from the contact interaction 
 Lagrangian 
 \cite{llmodel,Taekoon} with the SM Lagrangian.
 We correct the  leading order (LO) cross section calculation
  for higher order QCD effects using a mass-dependent
 $K$-factor. 
The $K$-factor is 
 defined as the ratio of the NNLO
  Drell-Yan cross section calculation from Ref.~\cite{dynnlo} 
 to our calculated LO Drell-Yan cross section. 
 Limits are set independently 
 for each separate channel of the contact-interaction 
 Lagrangian: $LL$, $LR$, $RL$ and $RR$,
 and $\eta_{ij} = \pm 1$. 
 The first letter indicates the helicity of the
 quark current and the second letter indicates the helicity of the lepton
 current. These terms are  strongly constrained by 
 atomic parity-violation measurements (APV)\cite{APV}, implying $\Lambda > 10$
 TeV. However, parity-conserving or other symmetric combinations of 
 these terms, such as $LL+RR$, $LR+RL$ \cite{KBabu,VBarger}, $LL-LR$, $RL-RR$ 
 \cite{ANelson},
 vector-vector $(VV=LL+RR+LR+RL)$ \cite{NBarto},
  and axial vector - axial vector
 $(AA=LL+RR-LR-RL)$
 \cite{NBarto,DWyler}, are not constrained by APV. Our
 measurements impose strong constraints on all of these models.   

 The limit on the quark-electron compositeness scale is calculated using a 
 Bayesian analysis of  the shape of the mass distribution of events.
 The expected number of events in the 
 $k^{\rm th}$ mass bin is denoted by 
 $N_{\Lambda}^k = b^k + {\cal L}{\epsilon^k} {\sigma_{\Lambda}^k}$, 
 where  ${\sigma_{\Lambda}^k}$
 is the predicted cross section including compositeness ($\Lambda \rightarrow
 \infty$ gives the SM cross section). To reduce the normalization
 uncertainty in the theory for the limit-setting analysis, 
 the SM prediction for the number of events in the
 $Z$ boson mass bin is normalized to the observed number of events. 
 The posterior probability density for the compositeness scale $\Lambda$, 
 given the observed data distribution $(D)$, is given by
\begin{eqnarray}
P({\Lambda}{\mid}{D})=
\frac{1}{A}
\int db \; {d\epsilon}
\prod_{k=1}^n 
\left[ 
\frac {{e^{-{N_{\Lambda}^k}}}{N_{\Lambda}^k}^{N_{0}^k}} {{N_{0}^k}!}
P({b^k},
{\epsilon^k})
\right] 
P({\Lambda}). \nonumber
\label{post_prob}
\end{eqnarray} 
  The bin-to-bin correlations in the value 
 and the uncertainty on the background  are taken into account. 
 The prior distribution $P({\Lambda})$ is 
 uniform in 1/${\Lambda^2}$, to represent a prior approximately
 uniform in cross section. The resulting
 posterior density $P({\Lambda}{\mid}{D})$ peaks
 at $1/{\Lambda^2}=0$ and falls monotonically with increasing 1/${\Lambda^2}$.
 The 95\% CL lower limit
  is defined by
  $ \int_{\Lambda}^{\infty} P({\Lambda^\prime}{\mid}{D}) d{\Lambda^\prime}
 =0.95$. 
 The limits for various 
 helicity options are shown in Table~\ref{lamda_model_limit}.

 In conclusion, we have measured the differential
 cross section for dielectron
 pair production at high dielectron mass. We find no significant deviation 
 from the
 SM. We have used the data to set limits on the
 quark-electron compositeness scale in the context of an effective 
 contact interaction. In the chiral channels
 (i.e., either the quark or the lepton current
 is not vector or axial-vector), 
 the 95\% CL lower limits on $\Lambda^+$ vary between
 3.3 and 
 4.2 TeV, and limits on  $\Lambda^-$ vary between 3.6 and
 5.1 TeV. The $VV$ and
 $AA$ limits are more stringent, varying between 4.7 and 
 4.9 TeV for $\Lambda^+$ 
 and 5.5 and 6.1 TeV for $\Lambda^-$. These are the best limits to date
 on quark-electron
 compositeness and are fairly independent of the helicity structure of
 the contact interaction. 
    
We thank T. Sj$\ddot{\rm o}$strand for discussions 
regarding {\small PYTHIA} and W. L. Van Neerven for providing the code 
 to compute the 
NNLO SM Drell-Yan cross section. 
We thank the staffs at Fermilab and collaborating institutions for their
contributions to this work, and acknowledge support from the 
Department of Energy and National Science Foundation (U.S.A.),  
Commissariat  \` a L'Energie Atomique (France), 
Ministry for Science and Technology and Ministry for Atomic 
   Energy (Russia),
CAPES and CNPq (Brazil),
Departments of Atomic Energy and Science and Education (India),
Colciencias (Colombia),
CONACyT (Mexico),
Ministry of Education and KOSEF (Korea),
and CONICET and UBACyT (Argentina).

\end{document}